\documentclass[10pt,conference,letterpaper]{IEEEtran}

\usepackage{amsmath,amsfonts,amsthm}
\usepackage{graphicx}
\usepackage{textcomp}
\usepackage{multirow}
\usepackage{subcaption}
\usepackage[ruled,linesnumbered,noend]{algorithm2e}
\usepackage{algpseudocode}
\usepackage{array}
\usepackage{multicol}
\usepackage{vwcol}
\usepackage{tikz}
\usepackage[most]{tcolorbox}
\usepackage{todonotes}
\usepackage{pifont}
\usepackage[figuresleft]{rotating}
\usepackage{color, colortbl}
\usepackage{listings}
\usepackage{amssymb}
\usepackage{url}
\definecolor{LightCyan}{rgb}{0.88,1,1}

\newtheorem{theorem}{Theorem}

\newtheorem{example}{Example}[section]

\newtcolorbox{mydefinition}[1][]{%
    enhanced,
    colback=blue!5!white,
    colframe=blue!75!black,
    fonttitle=\bfseries,
    title={Informative Metric Subset Problem},
    #1
}

\title{Metric Criticality Identification for Cloud Microservices}

\author{
\IEEEauthorblockN{
Akanksha Singal\IEEEauthorrefmark{1}\IEEEauthorrefmark{2},
Divya Pathak\IEEEauthorrefmark{1},
Kaustabha Ray\IEEEauthorrefmark{1},
Felix George\IEEEauthorrefmark{1},
Mudit Verma\IEEEauthorrefmark{1},
Pratibha Moogi\IEEEauthorrefmark{1}
}
\IEEEauthorblockA{\IEEEauthorrefmark{1}IBM Research - India}
\IEEEauthorblockA{\IEEEauthorrefmark{2}IIIT Delhi}
}

\begin{document}

\maketitle

\begin{abstract}
Modern cloud-native applications built on microservice architectures present unprecedented challenges for system monitoring and alerting. Site Reliability Engineers (SREs) face the daunting challenge of defining effective monitoring strategies across multitude of metrics to ensure system reliability, a task that traditionally requires extensive manual expertise. The distributed nature of microservices, characterized by stochastic execution patterns and intricate inter-service dependencies, renders the traditional manual approach of navigating the vast metrics landscape computationally and operationally prohibitive. To address this critical challenge, we propose KIMetrix, a data-driven system that automatically identifies minimal yet comprehensive metric subsets to aid SREs in monitoring microservice applications. KIMetrix leverages information-theoretic measures, specifically entropy and mutual information, to quantify metric criticality while considering the stochastic execution patterns inherent in microservice topologies. Our approach operates solely on lightweight metrics and traces, eliminating the need for expensive processing of unstructured logs, and requires no expert-defined training data. Experimental evaluation on state-of-the-art real-world microservice benchmark datasets demonstrates KIMetrix's effectiveness in identifying critical metric subsets that provide comprehensive system coverage while significantly reducing the burden on SREs. By automating the identification of essential metrics for alerting, KIMetrix enables more reliable system monitoring without overwhelming operators with false positives or missing critical system events.
\end{abstract}


\section{Introduction}
\label{sec:introduction}

\noindent
The surge in the complexity of modern applications has led to a substantial increase in the volume of observability data \cite{monitorassistant, msr, thalheim2017sieve}. This influx of data is further amplified by the widespread adoption of application deployment on distributed cloud platforms, where observability becomes paramount for understanding the health and performance of these intricate systems \cite{alerts4,Chen2019,gan2019open}. 
Observability data is multi-modal, each characterizing different facets of the underlying system \cite{alerts4,Chen2019,Chen2020,gan2019open}. For example, logs provide detailed records of events and activities within each service, offering valuable insights into the behavior of the system, traces provide records of paths governing path flows by individual executions of application invocations while metrics offer quantitative measurements of various aspects such as resource utilization, response times, and error rates, throughput, etc., aiding in performance analysis and dynamic resource management. 

Site Reliability Engineers (SREs) rely on metrics to define alerts providing timely notifications when a system's behaviour deviates from the expected performance. Metrics provide light-weight, real-time, quantitative view of the system's performance, health, and behavior critical towards production systems uptime. Defining alerts on metrics often necessitates manual subject matter expertise with white-box know-how on the intricate operations of each specific application and its implications on the environment it is deployed on. Such manual intervention, whilst being expensive, also becomes infeasible with thousands of observability metrics at play. An overly conservative strategy with alerts on a large number of metrics can lead to a large number of false positives, whilst a limited set of alerts can potentially miss out on critical events. Defining alerts on the critical metrics that ensure thorough coverage is thus an extremely challenging task. Automated techniques have been explored in literature to alleviate such scenarios, however, most of such techniques focus on traditional monolithic applications. 


Recently, there has been a significant shift towards the microservice architecture, particularly in cloud-native environments wherein applications comprise smaller, independent units that interact with each other, modeled as Directed Acyclic Graphs (DAGs), with vertices represent microservices and edges their interactions \cite{gan2019open}. This architecture offers benefits like scalability, flexibility, and the ability to deploy updates independently. However, traditional monitoring tools are ill-equipped to handle the nuances of microservices, which produce large volumes of observability data, far exceeding that of monolithic systems, making it challenging for SREs to define appropriate alerts. 
The problem of runtime alerting in microservices context 
is much more complex than the one for their monolithic counterparts due to: 1) the large number of underlying microservices, 2) the complex call relationships between them, and 3) the stochastic nature of executions governing the call relationships.

While there has been substantial work on metrics and the subsequent alert definitions for monoliths, most of these do not consider the intricacies introduced by microservice applications.
Only a small handful of recent proposals~\cite{cinque2022micro2vec, panahandeh2024serviceanomaly}, to the best of our knowledge in recent literature, have focused on the microservice context. However, these approaches do not consider the significance of pivoting an informative set of metrics toward aiding Subject Matter Experts to defining potential alerts.  
To address this, we propose the Informative Metric Subset Problem and introduce KIMetrix, that recommends critical metric set to aid SREs in defining alerts without without requiring expert training data. KIMetrix operates solely on metrics and traces, bypassing the need for high overhead processing of unstructured logs. In summary, the main contributions of this paper are:

\begin{itemize}
    \item We leverage a combination of entropy with mutual information to quantify the relative importance of metrics and prove NP-Completeness of the informative metric subset problem
    \item We then present KIMetrix to identify an approximate minimal set of metrics for each microservice by considering the stochastic nature of invocation of microservices exploiting the DAG topology
    \item We demonstrate the effectiveness of our approach on state-of-the-art real world benchmark microservice datasets
\end{itemize}

\noindent
The rest of this paper is organized as follows. Section \ref{sec:problem_definition} presents the problem definition. Section \ref{sec:detailed_methodology} details our methodology. Section \ref{sec:system_design} presents the system architecture. Section \ref{sec:experimentsanddiscussion} presents experimental results. Section \ref{sec:related_work} discusses related work, and finally, Section \ref{sec:conclusion} concludes the work and indicates future avenues.

\begin{table}[h!]
    \centering
    \caption{Table of Notations}
    \scalebox{1}{
    \begin{tabular}{|c|c|}
        \hline
        \textbf{Notation} & \textbf{Description} \\
        \hline
        $M$ & Total Set of Microservices \\
        $m$ & Microservice \\
        $\pi$ & Pivot set \\
        $\epsilon$ & Weighted Threshold \\
        $\mathcal{A}(M, E)$ & Application topology \\
        $\sigma$ & Set of all metrics available for $\mathcal{A}$ \\
        $\sigma_m$ & Subset of metrics for $m$ \\
        $H(\Psi_m)$ & Set of entropy values for each metric in metric set $\Psi_m$ \\
        $\Psi_m$ & Metric set for microservice $m$ \\
        $\Psi_m^H$ & Sorted metric set by entropy \\
        $\sigma_m$ & Selected metric subset for $m$ \\
        $\mathcal{M}(\psi', \psi)$ & Mutual information between metrics \\
        $\Gamma_{\mathcal{A}}^*$ & Minimal subset of metrics for $\mathcal{A}$ \\
        $\mathcal{T}$ & Topological sorted set of $\mathcal{A}$ \\
        $\Phi$ & Set of microservices with a lower topological level \\
        $\Gamma(m)$ & Mapping between microservices and selected metrics \\
        $\sigma(m)$ & Subset of metrics selected for $m$ \\
        $\chi$ & Size of subset threshold \\
        $\rho$ & Path from the root in the topology \\
        $\tau$ & Tolerance \\
        $\alpha$ & Multiplicative factor \\
        $\beta$ & Additive factor \\
        $\eta$ & Number of iterations \\
        $\xi$ & Size of selected metric subset \\
        $\theta$ & Correlation Threshold \\
    \hline
    \end{tabular}
    }
\end{table}

\section{Problem Definition}
\label{sec:problem_definition}

\noindent
In this section, we formally define the problem addressed in this work. We begin by introducing the key notations.


Let $\mathcal{A}(M, E)$ denote a distributed microservice-based application, where:
\begin{itemize}
    \item $M = \{m_1, m_2, \dots, m_n\}$ is the set of microservices in the application.
    \item $E \subseteq M \times M$ is the set of directed edges representing communication between microservices. If $(m_i, m_j) \in E$, then $m_i$ communicates with $m_j$.
    \item $\Psi_m = \{\psi_1, \psi_2, \dots, \psi_k\}$ be the set of observable metrics associated with $m$.
    \item The total set of observable metrics for $\mathcal{A}$ is represented by $\sigma$ and $\mathcal{P}(\Psi_m)$ denotes the power set of $\Psi_m$
    \item  $H(\Psi_m)$ is the entropy set for each of the set of valuations of each metric that measures the uncertainty or amount of information contained in the metric
    \item $\mathcal{M}(\psi_1, \psi_2)$ is the mutual information between two arbitrary pair of metrics $\psi_1, \psi_2 \in \sigma$
\end{itemize}



\noindent
Let $P(\psi_k=\psi_k^i)$ denote the probability that the metric $\psi_k$ takes the value $\psi_k^i$ (binned in discrete intervals, since metrics are continuous valued). The entropy of $\psi_k$ is then defined as $H(\psi_k) = - \sum_{\psi_k^i} P(\psi_k = \psi_k^i) \log_2 P(\psi_k = \psi_k^i)$, where the summation is over all possible values of metric $\psi_k$. The mutual information between two metrics $\psi_1, \psi_2$  is defined as $\mathcal{M}(\psi_1, \psi_2) = \sum_{\psi_1, \psi_2} P(\psi_1, \psi_2) \log \left( \frac{P(\psi_1, \psi_2)}{P(\psi_1) P(\psi_2)} \right) $.
Mutual information represents a measure of redundancy between metrics. High mutual information indicates that knowing one of the metrics reduces uncertainty about the other, implying that the metrics are redundant in terms of the information they provide. Conversely, low mutual information means that the metrics provide largely independent information. Mutual information captures both linear and non-linear dependencies between metrics, making it ideal for complex microservice systems where metric relationships are often non-straightforward. We thus leverage mutual information to model the metric interplay in microservice applications. For each subset of metrics $\sigma^{+} \subseteq \sigma$, the aggregate mutual information is defined as $\sum_{\psi_1, \psi_2 \in \in \sigma ^{+}} \mathcal{M}(\psi_1, \psi_2)$. 
The informative metric subset problem is now formally defined as follows: 
\begin{mydefinition}
Given a microservice application $\mathcal{A}$ with microservices $M$, metrics $\Psi_m$ for each $m \in M$,  find a mapping $\Gamma_{\mathcal{A}}^* : M \rightarrow \mathcal{P}(\Psi_m)$ such that:

$$\Gamma_{\mathcal{A}}^* = \text{argmax}_{\Gamma_{\mathcal{A}}} \frac{\sum_{m \in M} \sum_{\psi \in \Gamma_{\mathcal{A}}(m)} H(\psi)}{\sum_{m \in M} |\Gamma_{\mathcal{A}}(m)|}$$

subject to:
\begin{enumerate}
    \item $\Gamma_{\mathcal{A}}(m) \subseteq \Psi_m$, $\forall m \in M$
    \item $\forall \psi_1 \neq \psi_2 \in \bigcup_{m \in M} \Gamma_{\mathcal{A}}(m):\ \mathcal{M}(\psi_1, \psi_2) \leq \epsilon$
    \item $\sum_{m \in M} |\Gamma_{\mathcal{A}}(m)| \leq \chi$
\end{enumerate}
\end{mydefinition}




\noindent
The informative metric subset problem aims at a balancing act by considering both informativeness (via entropy $H(\psi)$) and diversity (via mutual information constraint).
The objective, maximizes the average entropy per selected metric, across all microservices promoting the selection of metrics that are more informative and exhibit greater variability. Maximizing the average entropy rather than the total entropy ensures that the selected metrics are individually informative, not just collectively numerous. This prevents the selection of many low-entropy (uninformative) metrics that could inflate the total entropy if only the sum were maximized thereby ensuring compact, high-quality metric subsets, enabling SREs to monitor fewer signals without sacrificing observability. To ensure diversity and reduce redundancy, the constraints 2 and 3 ensure the mutual information between any pair of selected metrics is bounded and enforces a global budget on the total number of metrics selected.

\begin{theorem}
Informative Metric Subset Problem is NP-Complete.
\end{theorem}

\noindent
We reduce from the decision version of the Maximum Weighted Clique problem, which asks: given an undirected graph $G = (V, \mathcal{L})$ with vertex weights $w: V \rightarrow \mathbb{R}^+$ and integers $\chi$ and $W$, does there exist a clique $C \subseteq V$ of size at most $\chi$ with total weight at least $W$? Construct an instance of the metric subset selection problem as follows: for each vertex $v \in V$, define a corresponding metric $\psi_v$ with entropy $H(\psi_v) = w(v)$. Set the mutual information function $\mathcal{M}(\psi_u, \psi_v)$ such that $\mathcal{M}(\psi_u, \psi_v) = 0$ if $(u,v) \in \mathcal{L}$ and $\mathcal{M}(\psi_u, \psi_v) > \epsilon$ otherwise. The decision version of our problem is then: Is there a subset $S \subseteq \{\psi_v \mid v \in V\}$ of at most $\chi$ metrics such that all pairwise mutual information is at most $\epsilon$ and $\sum_{\psi \in S} H(\psi) \geq W$? Such a subset corresponds exactly to a clique of size at most $k$ and total weight at least $W$ in $G$. Further, given the subset, we can verify in polynomial time if the subset weight (entropy) is at least $W$.  Hence, the decision version of the informative metric subset selection problem is NP-complete. $\blacksquare$ \\

\noindent
In practice, pre-specifying the value of $\chi$ and $\epsilon$ presents a significant challenge for SREs. These values are inherently data-dependent and varies dramatically across different microservice applications, system scales, and operational contexts, and can drift dramatically as systems evolve overtime with as applications scale, new microservices are deployed, or system behavior patterns change, making any initial estimate quickly obsolete. This uncertainty necessitates an automated approach that can automatically find the informative metric subset by analyzing the underlying data characteristics, metric information content, and correlation patterns, thereby eliminating the need for SREs to specify $\chi$ and $\epsilon$ explicitly and ensuring that the selected metric subset adapts to the inherent properties of the observability data.
In the subsequent section, we discuss the details of KIMetrix, that automatically identifies $\Gamma^{*}_{\mathcal{A}}$, with only a seed value of $\epsilon$, that is dynamically adjusted at runtime, without the necessity of specifying $\chi$. 

\section{Detailed Methodology}
\label{sec:detailed_methodology}
\noindent
We now introduce a suite of algorithms that KIMetrix comprises: \textbf{Algorithm 1} selects a subset of metrics for each microservice based on entropy and mutual information, ensuring that only the most informative metrics are retained. \textbf{Algorithm 2} extends this subset selection to account for the topological dependencies between microservices, refining the metric selection process using a topology-aware approach. \textbf{Algorithm 3} leverages the above algorithms in an iterative process with Additive Increase Multiplicative Decrease (AIMD) to dynamically adjust the Mutual Information threshold for metric selection, aiming to converge towards an approximate minimal subset. 
In the following subsections, we discuss our methodology in detail.


    \begin{algorithm}
        \caption{Microservice Metric Subset Selection}
        \label{alg
        }
        \KwIn{$m$: Microservice, $\pi$: Pivot set, $\epsilon$: Weighted Threshold}
        \KwOut{$\sigma_m$: subset of metrics for $m$}
        
        $H(\Psi_m) \gets $ compute entropy for each $\psi \in \Psi_m$\;
        
        $\Psi_m^H \gets \{ \psi \in \Psi_m \mid H(\psi) > 0 \}$ \; 
        
        $\Psi_m^H \gets \text{Sort } \Psi_m \text{ by entropy in non-increasing order}$\;
        
        \eIf{$\pi = \emptyset$}{
            $\sigma_m \gets \{\texttt{argmax}_{\psi} \Psi_m^H \}$ \; 
        }{
            $\sigma_m \gets \emptyset$ \; 
        }
        
        $\Psi_m^C \gets \Psi^{H}_m \setminus \sigma_m$ \; 
        
        \ForEach{$\psi \in \Psi_m^C$}{
            $\texttt{add\_metric} \gets \text{True}$\;
            
            \ForEach{$\psi' \in \sigma_m \cup \pi$}{
                $\mathcal{M}(\psi',\psi) \gets $ \texttt{MutualInformation} ($\psi'$, $\psi$)\;
                
                \If{$\mathcal{M}(\psi', \psi) > \epsilon$}{
                    $\texttt{add\_metric} \gets \text{False}$\;
                    \textbf{break}\;
                }
            }
            
            \If{$\textnormal{\texttt{add\_metric}}$}{
                $\sigma_m \gets \sigma_m \cup \psi$\;
            }
        }
        
        \Return $\sigma_m$\;
    \end{algorithm}

\subsection{Microservice Metric Subset Selection}

\noindent
For a microservice $m$, Algorithm~1 selects a metric subset using entropy and mutual information. Entropy ranks metrics by individual informativeness, while mutual information—derived from entropy—captures both linear and nonlinear redundancy, enabling consistent selection of compact, informative subsets. 
It first computes the entropy $H(\Psi_m)$ sorted in non-increasing order of their entropy values (Line 3) allowing prioritization metrics with the highest information content.
 If $\pi = \emptyset$, the metric with the highest entropy is chosen as the initial selected metric (Line 5). If $\pi$ is not empty, $\sigma_m$ is initialized as an empty set (Line 7), indicating that the algorithm will build the subset based on the pivot set. The algorithm constructs candidate metrics $\Psi_m^C$ by excluding the initially selected metric and ranks them by entropy to guide informed, non-redundant inclusion into the subset.
The core of the algorithm is an iterative process (Lines 9 - 17) where for each candidate metric, the algorithm checks its mutual information $\mathcal{M}(\psi', \psi)$ with all metrics already in $\sigma_m$ and the pivot set $\pi$. If the mutual information exceeds the threshold $\epsilon$, the metric is considered redundant and is not added to $\sigma_m$ (Line 14). Otherwise, the metric is added to the selected subset (Line 17). Finally, after evaluating all candidate metrics, the algorithm returns the subset $\sigma_m$ (Line 18) associated with the microservice $m$. 
Algorithm 1 greedily maximizes information while minimizing redundancy. We next introduce Algorithm 2, which extends this approach by incorporating microservice topology into the selection process

    \begin{algorithm}
    \caption{Topology Aware Subset Selection}
    \label{alg:topology_aware_subset_selection}
    \KwIn{$\mathcal{A}(M, E)$, Threshold $\epsilon$}
    \KwOut{$\Gamma : M \rightarrow \mathcal{P}(\sigma)$}
    $\mathcal{T} \leftarrow$  Topology Sort  $\mathcal{A}$\;
    $\Gamma(\text{Root}(\mathcal{T})) \leftarrow$ Algorithm 1 ( Root($\mathcal{T}$), $\emptyset$, $\epsilon$ ) \;
    $\mathcal{M} \gets \mathcal{T} \setminus \{\text{Root}(\mathcal{T})\}$

    \ForEach{ $m \in \mathcal{M}$ }{
        $\Phi \gets \{m' \in \mathcal{M} \mid L(m') < L(m)\}$\;
        \Comment{$L(m)$ is topology sort level of $m$} \;
        $\pi_m \gets \bigcup_{m' \in \Phi} \Gamma(m')$\;
        \Comment{$\Gamma(m)$ is subset of metrics selected for $m$} \;
        $\Gamma_m \gets \emptyset$\;

        \ForEach{\textnormal{path} $\rho$ \textnormal{from} $\text{Root}(\mathcal{T})$ to $m$}{
            $\epsilon \gets \epsilon \times \frac{1}{\mathbb{P}(\rho)}$\;
            $\Gamma_m \gets \Gamma_m \cup \text{Algorithm 1}(m, \pi_m, \epsilon)$\;
        }

        $\Gamma(m) \gets \Gamma_m$\;
    }
    \Return{$\Gamma$}
    \end{algorithm}

\begin{figure*}[h!]
    \centering
    \begin{minipage}{0.49\textwidth}
        \centering
        \includegraphics[scale = 0.64]{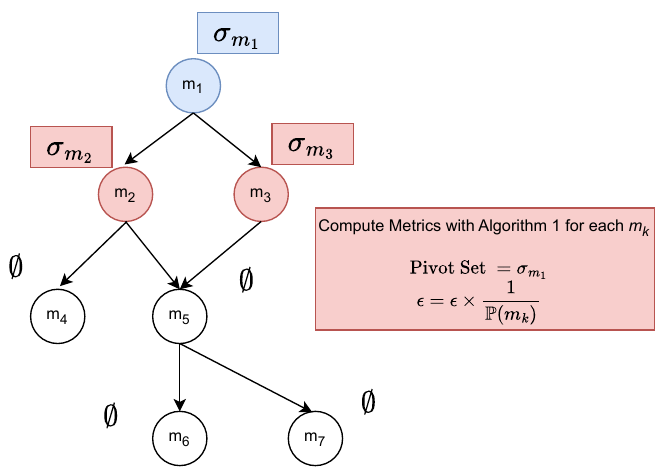}
        \subcaption{Root and 1-Successor Computations\label{fig:3}}
    \end{minipage}%
    \hfill
    \begin{minipage}{0.49\textwidth}
        \centering
        \includegraphics[scale = 0.64]{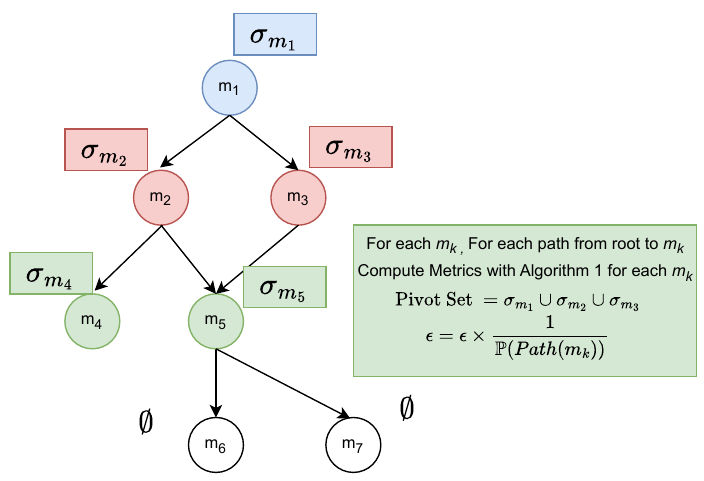}
        \subcaption{Singular Path Microservices\label{fig:4}}
    \end{minipage}

    \begin{minipage}{0.5\textwidth}
        \centering
        \includegraphics[scale = 0.6]{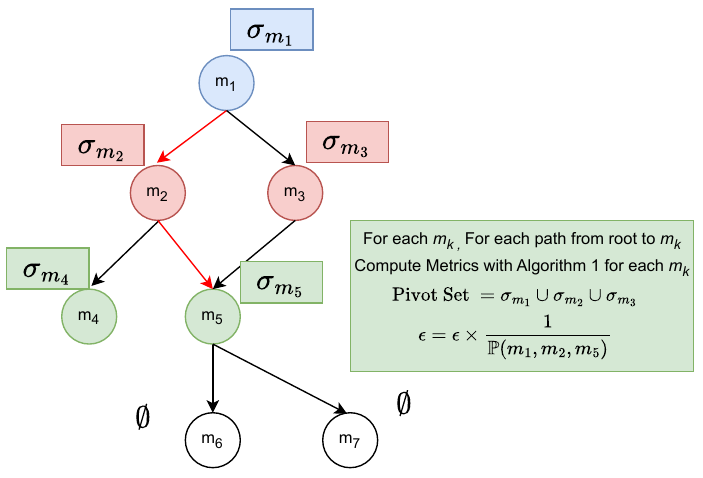}
        \subcaption{Multi-Path Microservice\label{fig:5}}
    \end{minipage}%
    \hfill
    \begin{minipage}{0.5\textwidth}
        \centering
        \includegraphics[scale = 0.6]{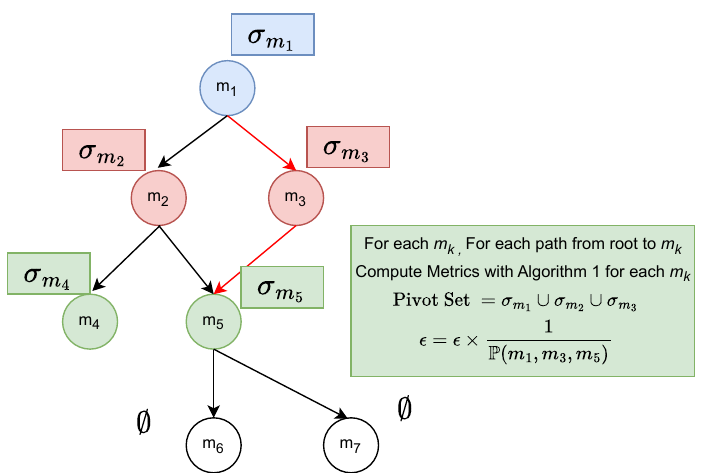}
        \subcaption{Multi-Path Microservice\label{fig:6}}
    \end{minipage}

    \caption{Execution Flow of Algorithm 2}
    \label{fig:algo2}
\end{figure*}

\subsection{Topology Aware Microservice Subset Selection} \label{sec:microservice}

\noindent
Algorithm \ref{alg:topology_aware_subset_selection} leverages Algorithm 1 to find a subset of metrics for \emph{each} microservice in a microservice-based application, considering the application's topology. The algorithm computes this subset by considering the topology of the application quantified in terms of probabilities of path executions. Thus, it exposes for each microservice the informative metric subset from a microservice-topology aware perspective.

The algorithm first performs topological sort on \( \mathcal{A}(M, E) \) to determine the order in which microservices should be processed (Line 1). It computes metrics for the root microservice using Algorithm 1 with an empty pivot set (Line 2), then iterates over the remaining microservices. For each microservice \( m \), the algorithm identifies its predecessor microservices, \( \Phi \), based on the topology sort:
    
    \[
    \Phi \gets \{m' \in \mathcal{M} \mid L(m') < L(m)\} \text{ (Line 5)}
    \]
\noindent
where \( L(m) \) represents the topology level of \( m \). The pivot set \( \pi_m \) is constructed by taking the union of the selected metric subsets \( \Gamma(m') \) of all predecessor microservices \( m' \in \Phi \):
    
    \[
    \pi_m \gets \bigcup_{m' \in \Phi} \Gamma(m') \text{ (Line 7)}
    \]    

\noindent
For each path from the root of the topology to microservice \( m \), the algorithm adjusts the threshold \( \epsilon \) based on the probability of the path (Line 11). Low-probability paths get higher thresholds (the inverse of the probability will be large), retaining critical metrics for rare fault scenarios. High-probability paths get lower thresholds (the inverse of the probability will be small), creating a balance between frequent and infrequent execution paths and not overly favoring one path over the other. We use traces to compute path probabilities. Let \( \mathbb{P}(m_i \mid m_{i-1}) \) denote the conditional probability of microservice \(m_i\) executing after \(m_{i-1}\), derived from traces. The path probability $\mathbb{P}(\rho)$ is then:

\[
\mathbb{P}(m_1 \rightarrow m_2 \rightarrow \dots \rightarrow m_n) = \mathbb{P}(m_1) \cdot \prod_{i=2}^{n} \mathbb{P}(m_i \mid m_{i-1})
\]
Next, the algorithm invokes Algorithm 1 to select a subset of metrics for \( m \), using the pivot set \( \pi_m \) and the adjusted threshold \( \epsilon \). The selected metric for the microservice,  \( \Gamma(m) \),  is the union of all the metrics returned for each path:
    
    \[
    \Gamma_m \gets \Gamma_m \cup \text{Algorithm 1}(m, \pi_m, \epsilon) \text{ (Line 12) }
    \]
    
\noindent
By weighting redundancy based on path probabilities, the algorithm captures both information overlap and execution likelihood, enabling more effective and context-aware metric selection in microservice-based systems.

\begin{example}
    
Figure \ref{fig:algo2} illustrates Algorithm 2's execution on a DAG with microservices $m_1, m_2, m_3, m_4, m_5, m_6, m_7$. 
The algorithm starts with topological sort, then processes root $m_1$ using Algorithm 1 with empty pivot $\pi = \emptyset$, producing subset $\sigma_{m_1}$. This becomes the pivot for children $m_2$ and $m_3$, which use adjusted thresholds $\epsilon \times \frac{1}{\text{Pr}(m_k)}$.

For $m_4$ (child of $m_2$), the pivot set is constructed with all paths from the root to the  \( m_4 \). In this scenario, there is only one path, \( m_1, m_2, m_4 \). The pivot set for \( m_4 \) is set to the union of the metric subsets from the previous nodes in the path, i.e., \( \pi_{m_4} = \sigma_{m_1} \cup \sigma_{m_2} \). Using Algorithm 1, the subset \( \sigma_{m_4} \) is computed with an adjusted threshold based on the probability of the path \( m_1, m_2, m_4 \).
For node \( m_5 \), the algorithm considers all paths leading to it from the root \( m_1 \) with the pivot \( \pi_{m_5} = \sigma_{m_1} \cup \sigma_{m_2} \cup \sigma_{m_3} \).  However, for \( m_5 \) there are two distinct paths from the root, \( m_1, m_2, m_5 \) and \( m_1, m_3, m_5 \). Algorithm 1 is then executed with this pivot set and an adjusted threshold \( \epsilon = \epsilon \times \frac{1}{\mathbb{P}(m_1, m_2, m_5)} \) and \( \epsilon = \epsilon \times \frac{1}{\mathbb{P}(m_1, m_3, m_5)} \) for each path. The resultant subset \( \sigma_{m_5} \) is the union of the metrics identified from each path.
Finally, $m_6$ and $m_7$ are processed similarly as children of $m_5$.

\end{example}

\noindent
Algorithm 2 exploits the hierarchical topology, refining metric subsets as pivots for downstream nodes. However, it depends on a manually set $\epsilon$ to control diversity and subset size. To address this, we propose an automated AIMD (Additive Increase Multiplicative Decrease)-based approach that eliminates the need for manual $\epsilon$ tuning. It adaptively explores the parameter space by leveraging correlations between selected and non-selected metrics to guide informative subset selection, alleviating the hurdles of SREs.

\begin{algorithm}
    \caption{AIMD Driven Informative Metric Subset }
    \label{alg}
    
    \KwIn{$\mathcal{A}(M,E)$ : \text{application}; $\tau$ : tolerance;
    $\alpha$ : \text{multiplicative factor};
    $\beta$ : \text{additive factor};
    $\eta$ : \text{iterations}; $\epsilon$ : \text{initial threshold}}
    \KwOut{$\Gamma_{\mathcal{A}}^* : M \rightarrow \mathcal{P}(\sigma)$}
    
    \BlankLine

    $\mathcal{E} \gets \texttt{correlation}(\mathcal{A}, \sigma)$ \;
    
    $\texttt{subset\_sizes} \gets \{ |\sigma| \}$\;
    $\texttt{subsets} \gets \emptyset $ \; 
    
    \For{$i = 1$ \KwTo $\eta$}{
        $\Gamma \gets \texttt{Algorithm2}(\mathcal{A}, \epsilon)$ \; 
        $\xi \gets |\Gamma|$ \;
        $\texttt{is\_within\_tolerance} \gets \texttt{True}$ \;
        $\texttt{subsets} \gets \texttt{subsets} \cup \{(\Gamma, \mathcal{C}(\Gamma))\}$ \; 
    
        \ForEach{$s \in \textnormal{\texttt{subset\_sizes}}$}{
            \If{$| \xi - s | \geq \tau$}{
                $\texttt{is\_within\_tolerance} \gets \texttt{False}$ \;
                \textbf{break} \;
            }
        }
        
        \If{$\textnormal{\texttt{is\_within\_tolerance}}$}{
            $\epsilon \gets \epsilon + \beta$ \;
        }\Else{
            $\epsilon \gets \epsilon \times \alpha$ \;
        }
        
        $\texttt{subset\_sizes} \gets \texttt{subset\_sizes} \cup \{\xi\}$ \;
    }
    
    

    $\mathcal{C}_{\texttt{max}} \gets$ \texttt{max}$(\{ c \mid (\Gamma, c) \in \texttt{subsets} \})$ \;

    $\Gamma_{\mathcal{C}_{max}} \gets \{ \Gamma \mid (\Gamma, c) \in \texttt{subsets},\ c = \mathcal{C}_{\texttt{max}} \}$ \;

    $\Gamma_{\mathcal{A}}^* \gets \texttt{argmin}_{\Gamma \in \Gamma_{\mathcal{C}_{max}}} |\Gamma|$ \;

    \Return $\Gamma_{\mathcal{A}}^*$ \;

\end{algorithm}

\subsection{Automated Informative Minimal Subset Construction} \label{sec:automatedminimal}

\noindent
Algorithm 3 uses an Additive Increase Multiplicative Decrease (AIMD) strategy to automatically tune the threshold \( \epsilon \) for informative metric subset selection. It initializes key parameters for the iterative process: (i) tolerance \( \tau \) for acceptable deviation in subset size across iterations, (ii) AIMD factors \( \alpha \) (multiplicative) and \( \beta \) (additive), (iii) number of iterations \( \eta \), and (iv) initial seed \( \epsilon \) to influence the first subset selection.

The algorithm first computes the correlation between each pair of metrics (Line 1), leveraging it as a proxy to quantify the informativeness of each subset selected in each iteration of the AIMD.
We consider a pair of metrics \( (\psi_i, \psi_j) \) with a correlation coefficient \( \Theta(\psi_i, \psi_j) \) above a threshold \( \theta \) as redundant. The presence of one of the correlated metrics in a subset \( \Gamma \) facilitates the ability to infer the other metric. Thus, the coverage of a subset \( \mathcal{C}(\Gamma) \) is then defined as:

\[
\mathcal{C}(\Gamma) = \frac{|\{\psi_k \in \sigma \mid \psi_k \in \Gamma \lor \exists \psi_m \in \Gamma: \Theta(\psi_k, \psi_m) > \theta\}|}{|\sigma|}
\]

\noindent
For unsupervised time series correlation, metric pairs are treated as correlated if classified as similar, eliminating the need for a fixed threshold \( \theta \).
KIMetrix supports any time series correlation measure (e.g., Kendall’s, Spearman’s), allowing SREs to plug in their preferred choice (Section~\ref{sec:experiments}).
The algorithm begins by initializing \texttt{subset\_sizes} with the original total number of metrics and an empty set \texttt{subsets}. It proceeds for \( \eta \) iterations (Lines 4--21), invoking Algorithm 2 to compute the metric subset \( \Gamma \) using the current threshold \( \epsilon \) (Line 5). The resulting subset size \( \xi \) plays a central role in the AIMD-based adjustment. 
Based on the tolerance check, if the current subset size \( \xi \) is within \( \tau \), \( \epsilon \) is increased additively by \( \beta \) (Line 14) to promote smaller subsets. Otherwise, \( \epsilon \) is decreased multiplicatively by \( \alpha \) (Line 16) to include more metrics and stabilize selection.

Once all iterations are finished, it first calculates the maximum coverage obtained in the iterations and finally returns the subset with the minimum size amongst all those sets with the maximum coverage (Lines 18-21). 
Algorithm 3 leverages the AIMD approach to iteratively refine the selection of metric subsets, balancing between under-selection and over-selection of metrics. By adjusting the threshold \( \epsilon \) based on the tolerance \( \tau \), KIMetrix produces an approximate minimalist set of metrics that an SRE can leverage to define alerts. Further, as we demonstrate in Section \ref{sec:experiments}, KIMetrix can also expose all the subsets in the AIMD iterations along with their coverage percentages allowing a system administrator to explore the trade-offs in subset size versus coverage. 

\subsection{Time Complexity}

\noindent
Let $\rho_{max}$ denote the maximum path length and $\Psi_{max}$ denote the maximum size of the set of metrics across all microservices. The worst case time complexity of KIMetrix is $\mathcal{O}(\eta \cdot |M| \cdot \rho_{max} \cdot (\Psi_{max})^2)$. In the next section, we discuss KIMetrix's system architecture. 

\section{System Architecture}
\label{sec:system_design}

\begin{table*}[h]
\centering
\scalebox{0.9}{
\begin{tabular}{|>{\centering\arraybackslash}m{3cm}|>{\centering\arraybackslash}m{2cm}|>
{\centering\arraybackslash}m{2.5cm}|>
{\centering\arraybackslash}m{3.2cm}|>
{\centering\arraybackslash}m{2.2cm}|>{\centering\arraybackslash}m{2.2cm}|}
\hline
\textbf{Correlation Methodology} & \textbf{Type of Data} & \textbf{Correlation Pruning Threshold} &\textbf{Subset Size, Epsilon ($k$ = $\mid M \mid$ , $\epsilon$)} & \textbf{Coverage Percentage $\mathcal{C}$} & \textbf{Coverage Percentage $\mathcal{C}_A$} \\ \hline
\multirow{2}{*}{Mutual Information} & Healthy & $0.04$ &  ($69, 7.5 \times 10^{-7}$)  & $76.98$\% & $75.00$\% \\ \cline{2-6} & Mix & $0.05$ & ($95$, $3 \times 10^{-2}$) & $86.90$\% & $76.85\%$ \\ \hline
\multirow{2}{*}{Pearson's R} & Healthy & $0.15$ & ($67$, $7.4\times 10^{-7}$)   & $74.20$\% & $73.15$\% \\ \cline{2-6} & Mix  & $0.27$ & ($74$, $2 \times 10^{-2}$)  & $78.96$\%  & $71.30$\% \\ \hline
\multirow{2}{*}{Spearman's Rank} & Healthy & $0.16$ & ($69$, $8 \times 10^{-7}$) & $74.07$\% & $72.22$\% \\ \cline{2-6} & Mix & $0.13$ & ($96$, $3 \times 10^{-2}$)  & $89.68$\% & $76.85$\% \\ \hline
\multirow{2}{*}{Kendall's Rank}       & Healthy & $0.12$ & ($67, 7.4\times 10^{-7} $)& $72.22$\% & $71.30$\%\\ \cline{2-6} & Mix  & $0.08$ &  ($65$, $2.0 \times 10^{-2}$)  & $84.50$\% & $73.14$\%\\ \hline
\end{tabular}
}
\caption{\centering Comparison of Different Correlation Methodologies by Data Type and Coverage Metrics for Topology based approach, Total metrics = $253$, $M$ = Final Selected Metric Subset, $\epsilon$ = Epsilon value}
\label{tab:results_correlation_methods}
\end{table*}

\begin{figure}
\centering
\includegraphics[scale=0.6, trim={0.7cm 0.7cm 0.7cm 0.7cm}, clip]{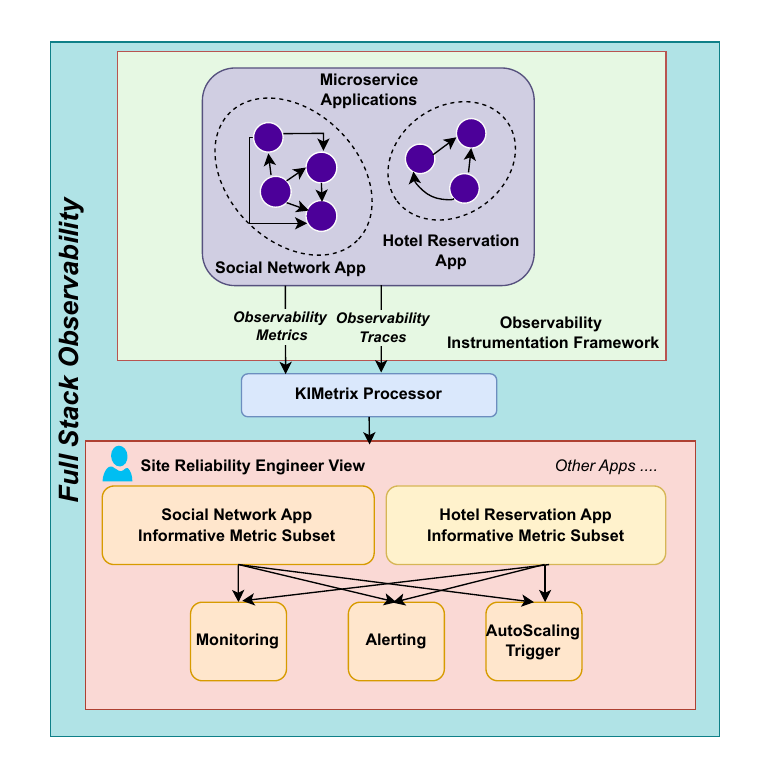}
\caption{System Architecture}
\label{fig:system_architecture}
\end{figure}

\noindent
In this section, we provide a high-level overview of the system architecture of KIMetrix, illustrated in Figure \ref{fig:system_architecture}. KIMetrix fits within any Full Stack Observability Framework exposing views of the Informative Metric Subset to aid SREs. For any microservice based application, KIMetrix relies on time series metrics, execution traces and the application topology that is obtained from the deployed Observability Stack Instrumentation layer. KIMetrix then utilizes the metrics coupled with the traces and the topology to compute the approximate minimal set of metrics. KIMetrix then exposes the informative metric subset to the SREs via the Observability Stack View endpoints. These subsets can then aid the SRE in different tasks such as monitoring, auto-scaling triggers to ensure SLA requirements. KIMetrix is inherently offline, in essence, necessary to execute only when the distribution of the workload changes to cope with variable workloads, KIMetrix can be executed at intervals to ensure the minimal subset remains up-to-date and reflective of current system behavior.

\section{Experiments and Discussion}
\label{sec:experimentsanddiscussion}

\subsection{Experimental Setup}

\noindent
In this section, we detail our experimental setup to evaluate KIMetrix. We use two benchmarks, one curated in-house on a live full-stack observability setup and another on a benchmark real-world dataset from DeathStarBench benchmark suite used in several microservice analytics.

For our in-house setup, we use the Quote of the Day (QoTD) microservice application \cite{rouf2024instantops} used in prior fault and root cause analysis studies. In order to evaluate KIMetrix for a myriad of real-world conditions, we leverage anomaly injection into the application. QoTD is hosted on an Openshift cluster \cite{RedHat2024}, an enterprise platform for containerized applications, with metrics and traces collected via Prometheus \cite{208870} and Instana \cite{instana}.
We generate real-world user behavior with a load generator that issues requests every 10–30 minutes across multiple instances, emulating large-scale concurrent access. Using the QoTD API, we cyclically inject anomalies into microservices for 1–10 minutes (sampled uniformly), followed by 10–30 minutes of healthy operation. We introduce 40 anomaly types affecting CPU, memory, latency, and error rates, enabling diverse and realistic conditions for evaluating subset selection and model robustness. Metrics and traces are collected every 3 seconds over one day, covering 253 unique metrics. We classify data as:
a) Healthy – No injected anomalies, variation due to load.
b) Mix – Includes periods with and without anomalies.

Additionally, we used the widely-used open-source DeathStarBench dataset \cite{gamma, gan2019open} for the Social Networking application, a representative microservice application, deployed on Kubernetes environment. The dataset contains approximately 40 million request traces and time series data for CPU, memory, I/O, and network metrics, with varying bottleneck intensities and durations to simulate real-world conditions, making bottleneck detection non-trivial. The dataset includes traces and metrics with multiple bottlenecks at different intensities that degrade the application performance but do not cause any faults or errors that can be trivially detected. VMs on hosts are induced with different types of interference (e.g., CPU and memory), resulting in the hosted microservices experiencing a mixture of interference patterns. 

\subsection{Results and Discussion} \label{sec:experiments}


\noindent

\noindent
Both the QoTD and DeathStarBench datasets comprise labels with respect to anomalous scenarios. Whilst KIMetrix is unsupervised and does not require this labeling, we leverage this labeling to evaluate KIMetrix compare it with other approaches. 
Let $\sigma_{A} \subseteq \sigma$ denote the set of time-series metric instances labeled with anomaly = 1. Then, we define the coverage for the anomalous scenarios as:

\[
\mathcal{C}_{A}(\Gamma) = \frac{|\{\psi_k \in \sigma_A \mid \psi_k \in \Gamma \lor \exists \psi_m \in \Gamma: \Theta(\psi_k, \psi_m) > \theta\}|}{|\sigma_A|}
\]

\noindent
Thus, $\mathcal{C}$ and $\mathcal{C}_A$ serve as effective measures of the usefulness of a subset of metrics to SREs that we utilize to evaluate KIMetrix. We now answer the following research questions: \\

\begin{figure}[t]
\centering
\includegraphics[width=0.45\textwidth]{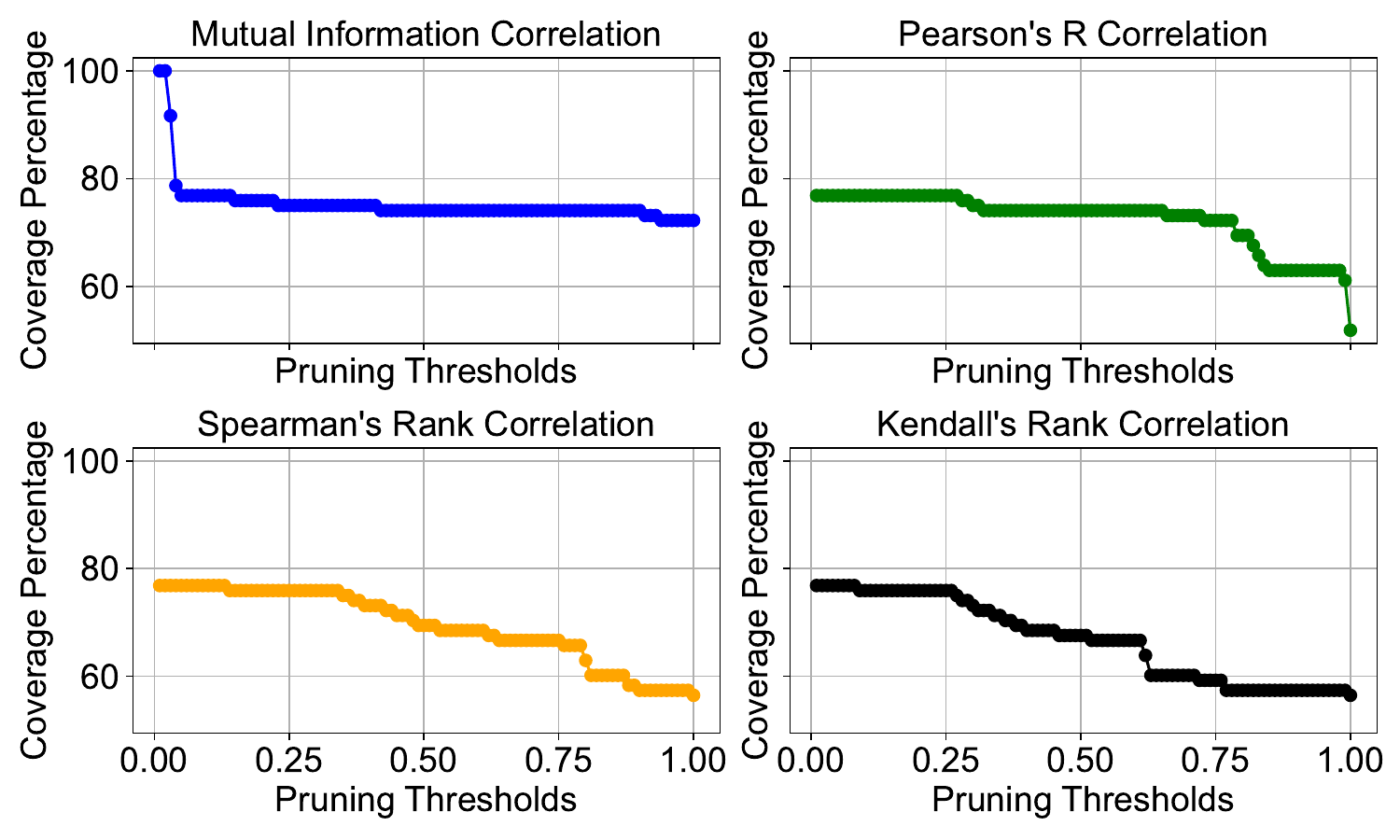} 
\caption{Comparing Pruning Thresholds and Coverage Across Correlation Methods for $\mathcal{C}_A$}
\label{fig:correlation-coverage}
\end{figure}

\begin{figure}[t]
\centering
\includegraphics[width=0.47\textwidth]{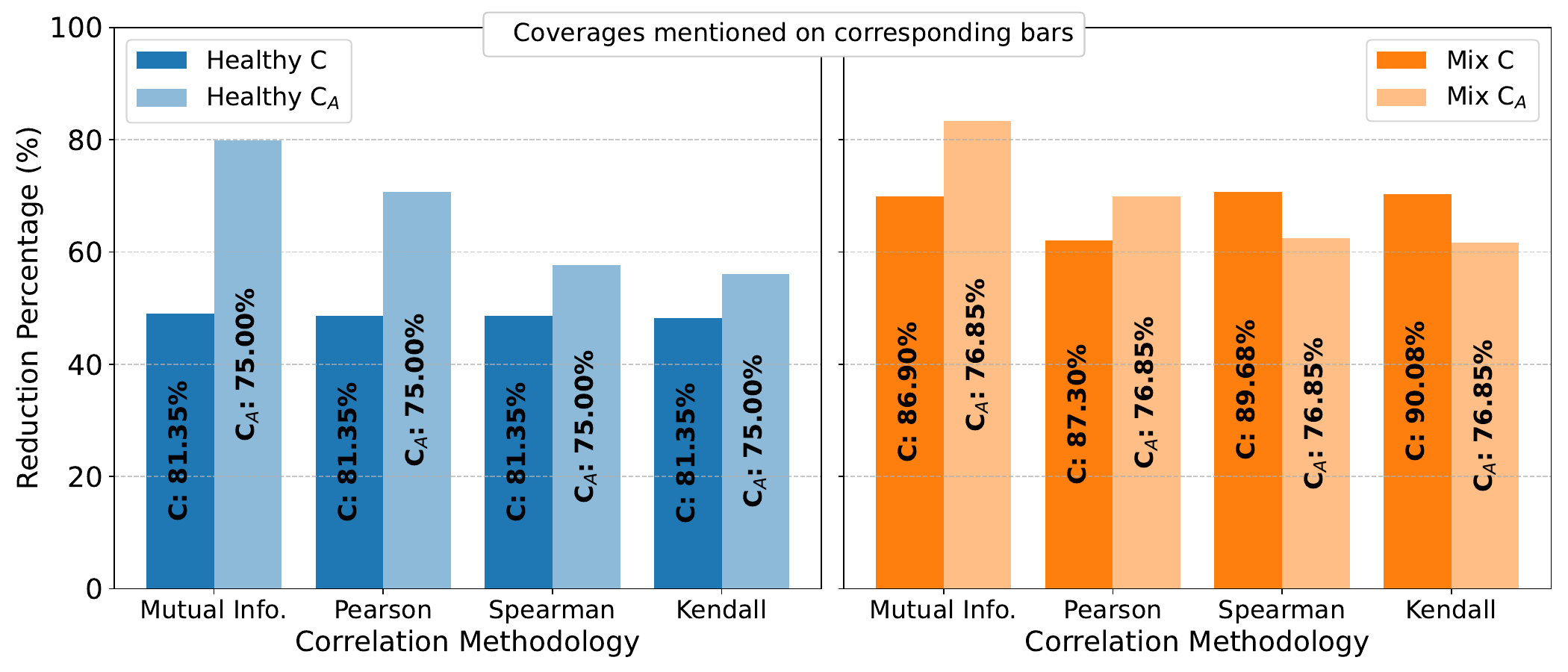} 
\caption{Reduction in the Metrics Space for QoTD}
\label{fig:reduction-plot}
\end{figure}

\noindent
\textbf{Q1: To what extent can we reduce the size of the metrics set without compromising coverage?}
To assess trade-offs across correlation measures, we evaluate Mutual Information, Pearson, Spearman, and Kendall correlations on both Healthy and Mix datasets. Although Mutual Information is not a traditional correlation metric, it effectively captures informativeness and can also be leveraged in this setting. 
Table~\ref{tab:results_correlation_methods}  gives an overview of subset sizes and epsilon values for various correlation measures with the listed correlation pruning thresholds on Healthy and Mix datasets for QoTD. Typically, overall coverage \( \mathcal{C} \) exceeds anomaly-specific coverage \( \mathcal{C}_A \), as it spans the entire metric space, while \( \mathcal{C}_A \) focuses on anomaly-affected metrics. By varying pruning thresholds from 0.0 to 1.0, Figure~\ref{fig:correlation-coverage} shows that KIMetrix consistently achieves high \( \mathcal{C}_A \) across all correlation types. Notably, Mutual Information supports more generalizable selection, balancing coverage and pruning. Figure~\ref{fig:reduction-plot} illustrates the corresponding metric reduction.

\begin{table}
\centering
\begin{tabular}{|c|c|c|}
\hline
\textbf{Type of Data} & \textbf{\shortstack{Subset \\ Selection }} & \textbf{\shortstack{AIMD $t_N$ \\ $N$ = $500$ (in hours)}} \\
\hline
\multirow{2}{*}{Healthy} & Flat & $12.00$ \\
\cline{2-3}
& Topology & $5.15$ \\
\hline
\multirow{2}{*}{Mix} & Flat & $6.90$ \\
\cline{2-3}
& Topology & $0.60$ \\
\hline
\end{tabular}
\caption{Execution Time comparison for Subset Selection Methods, $N$ = Total number of Iterations}
\label{tab:performance_metrics}
\end{table}

\noindent
\textbf{Q2: What additional benefits do we gain from incorporating topology of the application?} 
Algorithm 2 essentially invokes Algorithm 1 iteratively in a topological sort manner iteratively growing the subset for each microservice. Note that Algorithm 1 can also be invoked independently of Algorithm 2, wherein Algorithm 1 essentially executes on the union of all the metrics for all microservices thereby rendering a flat view of the application topology.
We now highlight the pivotal role of Algorithm 2 in KIMetrix.
For this experiment, we execute Algorithm 3 with 0.5 as the initial epsilon set for multiplicative and additive factors to be 0.4 and 0.005 on the QoTD dataset. Table \ref{tab:performance_metrics} highlights the reduction in execution time obtained by the topology-aware mechanism over a flat approach. The topology aware approach significantly reduces the search space at each level of the topology by expanding or contracting the subset selection based on path probabilities, which accelerates the convergence and enhances the overall efficiency. This results in drastically faster execution.

\begin{table*}[ht]\scalebox{0.9}{
\begin{tabular}{|c|c|c|c|c|c|c|c|c|c|c|c|}
\hline
\multirow{3}{*}{\textbf{\begin{tabular}[c]{@{}c@{}}Subset \\ Selection \\ Method\end{tabular}}} & \multirow{3}{*}{\textbf{\begin{tabular}[c]{@{}c@{}}Automatic \\ Subset \\ Return\end{tabular}}} & \multirow{3}{*}{\textbf{\begin{tabular}[c]{@{}c@{}}Subset \\ Size (K)\end{tabular}}} & \multicolumn{8}{c|}{\textbf{Correlation Methods}} & \multirow{3}{*}{\textbf{\begin{tabular}[c]{@{}c@{}}Execution \\ Time \\ (secs)\end{tabular}}} \\ \cline{4-11}
 & & & \multicolumn{2}{c|}{\textbf{Mutual Information}} & \multicolumn{2}{c|}{\textbf{Pearson's}} & \multicolumn{2}{c|}{\textbf{Spearman's Rank}} & \multicolumn{2}{c|}{\textbf{Kendall's Rank}} & \\ \cline{4-11}
 & & & \(\mathcal{C}_A\) & \(\mathcal{C}\) & \(\mathcal{C}_A\) & \(\mathcal{C}\) & \(\mathcal{C}_A\) & \(\mathcal{C}\) & \(\mathcal{C}_A\) & \(\mathcal{C}\) & \\ \hline
$\textit{SelectKBest}\textsuperscript{+}$ \cite{scikit-learn-selectkbest} & \ding{55} & 60 & 73.15\% & 63.88\% & 64.81\% & 68.65\% & \textbf{74.07}\% & 60.32\% & \textbf{74.07}\% & 60.32\% & 0.04 \\ \hline
$\textit{mRMR}\textsuperscript{+}$ \cite{mrmr} & \ding{55} & 60 & 73.15\% & 63.88\% & 66.67\% & 68.65\% & \textbf{74.07}\% & 61.11\% & \textbf{74.07}\% & 61.5\% & 39.30 \\ \hline
$\textit{Boruta}\textsuperscript{+}$ \cite{boruta} & \ding{51} & 26 & 71.30\% & 48.01\% & 62.96\% & 41.67\% & 64.81\% & 41.26\%  & 65.74\% & 43.25\%  & 57.50 \\ \hline
\textit{Max Weighted Clique} \cite{max-weighted-clique} & \ding{51} & $89, 68, 78, 85 \textsuperscript{*}$ & 74.07\% & 59.92\% & 65.74\% & 59.52\% & \textbf{74.07}\% &  55.15\% & \textbf{74.07}\% & 56.34\%  & 300.20 \\ \hline
\textit{\textbf{KIMetrix}} & \ding{51} & $\textbf{76,97,74,75}\textsuperscript{*}$ & \textbf{76.85\%} & \textbf{86.90\%}  & \textbf{76.85\%} & \textbf{87.30\%} & 73.15\% & \textbf{89.68\%}  & 73.15\% & \textbf{90.08\%} & \textbf{2080} \\ \hline
\end{tabular}
}
\caption{Comparison of Feature Selection Methods in the literature with KIMetrix on QoTD \textit{Mix} data with CPU and Memory metrics 
* Subset sizes for each correlation method, + Supervised methods, requires labelling}
\label{tab:feature_selection_comparison1}
\end{table*}

\begin{table*}[ht]\scalebox{0.87}{
\begin{tabular}{|c|c|c|c|c|c|c|c|c|c|c|c|}
\hline
\multirow{3}{*}{\textbf{\begin{tabular}[c]{@{}c@{}}Subset \\ Selection \\ Method\end{tabular}}} & \multirow{3}{*}{\textbf{\begin{tabular}[c]{@{}c@{}}Automatic \\ Subset \\ Return\end{tabular}}} & \multirow{3}{*}{\textbf{\begin{tabular}[c]{@{}c@{}}Subset \\ Size (K)\end{tabular}}} & \multicolumn{8}{c|}{\textbf{Correlation Methods}} & \multirow{3}{*}{\textbf{\begin{tabular}[c]{@{}c@{}}Execution \\ Time \\ (secs)\end{tabular}}} \\ \cline{4-11}
 & & & \multicolumn{2}{c|}{\textbf{Mutual Information}} & \multicolumn{2}{c|}{\textbf{Pearson's}} & \multicolumn{2}{c|}{\textbf{Spearman's Rank}} & \multicolumn{2}{c|}{\textbf{Kendall's Rank}} & \\ \cline{4-11}
 & & & \(\mathcal{C}_A\) & \(\mathcal{C}\) & \(\mathcal{C}_A\) & \(\mathcal{C}\) & \(\mathcal{C}_A\) & \(\mathcal{C}\) & \(\mathcal{C}_A\) & \(\mathcal{C}\) & \\ \hline
$\textit{SelectKBest}\textsuperscript{+}$ \cite{scikit-learn-selectkbest} & \ding{55} & 60 & 82.77\%& 83.33\%& 81.11\%& 81.11\%& 81.66\%& 81.66\%& 81.66\%
& 82.22\%& 0.29\\ \hline
$\textit{mRMR}\textsuperscript{+}$ \cite{mrmr} & \ding{55} & 60 & 95.55\%& 96.11\%& 88.88\%& 89.44\%& 85.0\%& 85.55\%& 85.0\%
& 85.55\%& 37.35\\ \hline
$\textit{Boruta}\textsuperscript{+}$ \cite{boruta} & \ding{51} & 149& 90.55\%& 88.88\%& 83.33\%& 83.33\%& 85.0\%& 85.0\%& 85.0\%
& 85.0\%
& 317.95\\ \hline
\textit{Max Weighted Clique} \cite{max-weighted-clique} & \ding{51} & $148, 142, 141, 141 \textsuperscript{*}$ & 82.77\%& 82.77\%& 71.66\%& 79.44\%& 72.22\%&  78.88\%& 72.22\%& 78.88\%& 4362\\ \hline
\textit{\textbf{KIMetrix}} & \ding{51} & $\textbf{114, 114, 114, 114\textsuperscript{*}}$& \textbf{99.44\%}& \textbf{99.44\%}& \textbf{98.33\%}& \textbf{98.89\%}& \textbf{98.89}\%& \textbf{99.44\%}& \textbf{98.89}\%& \textbf{99.44\%}& \textbf{5880}\\ \hline
\end{tabular}
}
\caption{Comparison of Feature Selection Methods in the literature with KIMetrix on \textit{DeathStarBench} CPU data\\
Total metrics = 180, * Subset sizes for each correlation method, + Supervised methods, requires labelling}
\label{tab:feature_selection_deathstar_bench_cpu}
\end{table*}

\begin{table*}[ht]\scalebox{0.89}{
\begin{tabular}{|c|c|c|c|c|c|c|c|c|c|c|c|}
\hline
\multirow{3}{*}{\textbf{\begin{tabular}[c]{@{}c@{}}Subset \\ Selection \\ Method\end{tabular}}} & \multirow{3}{*}{\textbf{\begin{tabular}[c]{@{}c@{}}Automatic \\ Subset \\ Return\end{tabular}}} & \multirow{3}{*}{\textbf{\begin{tabular}[c]{@{}c@{}}Subset \\ Size (K)\end{tabular}}} & \multicolumn{8}{c|}{\textbf{Correlation Methods}} & \multirow{3}{*}{\textbf{\begin{tabular}[c]{@{}c@{}}Execution \\ Time \\ (secs)\end{tabular}}} \\ \cline{4-11}
 & & & \multicolumn{2}{c|}{\textbf{Mutual Information}} & \multicolumn{2}{c|}{\textbf{Pearson's}} & \multicolumn{2}{c|}{\textbf{Spearman's Rank}} & \multicolumn{2}{c|}{\textbf{Kendall's Rank}} & \\ \cline{4-11}
 & & & \(\mathcal{C}_A\) & \(\mathcal{C}\) & \(\mathcal{C}_A\) & \(\mathcal{C}\) & \(\mathcal{C}_A\) & \(\mathcal{C}\) & \(\mathcal{C}_A\) & \(\mathcal{C}\) & \\ \hline
$\textit{SelectKBest}\textsuperscript{+}$ \cite{scikit-learn-selectkbest} & \ding{55} & 60 & \textbf{48.33\%}& \textbf{49.44\%}& 46.11\%& 46.11\%& \textbf{48.33\%}& \textbf{48.33}\%& \textbf{48.33\%}
& \textbf{48.33\%}& 0.05\\ \hline
$\textit{mRMR}\textsuperscript{+}$ \cite{mrmr} & \ding{55} & 60 & \textbf{48.33\%}& 47.77\%& 45\%& 45\%& 47.22\%& 47.22\%& 47.22\%
& 47.22\%& 12.77\\ \hline
$\textit{Boruta}\textsuperscript{+}$ \cite{boruta} & \ding{51} & 63& \textbf{48.33\%}& 48.33\%& \textbf{47.77\%}& \textbf{47.77\%}& \textbf{48.33\%}& \textbf{48.33\%}& \textbf{48.33}\%
& \textbf{48.33\%}& 116.23\\ \hline
\textit{Max Weighted Clique} \cite{max-weighted-clique} & \ding{51} & $77, 78, 55, 77\textsuperscript{*}$& 41.11\%& 43.33\%& 31.11\%& 31.11\%& 31.11\%& 43.33\%& 31.66\%& 43.88\%& 329\\ \hline
\textit{\textbf{KIMetrix}} & \ding{51} & $\textbf{45, 45, 46, 45\textsuperscript{*}}$& \textbf{48.33\%}& 48.33\%& 47.22\%& 47.22\%& \textbf{48.33\%}& \textbf{48.33\%}& \textbf{48.33\%}& \textbf{48.33\%}& \textbf{2617.74}\\ \hline
\end{tabular}
}
\caption{Comparison of Feature Selection Methods in the literature with KIMetrix on \textit{DeathStarBench} Memory data\\
Total metrics = 180, * Subset sizes for each correlation method, + Supervised methods, requires labelling}
\label{tab:feature_selection_deathstar_bench_mem}
\end{table*}

\noindent
\textbf{Q3: How does KIMetrix compare to established feature selection approaches in the literature?}
Table \ref{tab:feature_selection_comparison1} offers an in-depth evaluation of feature selection methods commonly used to remove redundant features / find subsets. SelectKBest and mRMR are well-known supervised feature selection methods and require a target feature to be specified as well as the subset size. We leverage the anomaly label as the target feature. Boruta, automatically selects subsets but also requires the target feature. We additionally compare with the Max Weighted Clique method by modeling the correlation between features as a weighted graph (nodes representing metrics, edges only on mutual information threshold ($\epsilon$) satisfaction and node weights as entropy of the metric) and selecting the maximum weighted clique as the optimal feature subset.

We first evaluate the coverage metric \( \mathcal{C}_A \). While SelectKBest and mRMR achieve moderately high \( \mathcal{C}_A \) (64.81\%–74.07\%), they require the subset size to be predefined. Boruta selects 26 metrics with slightly lower coverage (62.96\%–71.30\%) and higher runtime due to its iterative nature. KIMetrix, in contrast, returns multiple subsets (sizes 76, 97, 74, and 75) with higher average coverage and smaller sizes than Max Weighted Clique. Although KIMetrix incurs a greater computation time, subset selection is infrequent and typically occurs during scheduled updates, rendering the offline cost negligible for SREs. For overall coverage \( \mathcal{C} \), KIMetrix significantly outperforms other methods. Supervised approaches like SelectKBest, mRMR, and Boruta favor labeled anomalies, achieving higher \( \mathcal{C}_A \) but lower \( \mathcal{C} \). Max Weighted Clique, though unsupervised, lacks microservice topology awareness. By incorporating this topology, KIMetrix ensures strong performance across both \( \mathcal{C}_A \) and \( \mathcal{C} \), making it effective even in unlabeled settings. Table \ref{tab:feature_selection_deathstar_bench_cpu} lists the same for the DeathStarBench CPU dataset wherein we observe KIMetrix obtaining significantly better coverage as opposed to other strategies with all the correlation measures considered. Table~\ref{tab:feature_selection_deathstar_bench_mem} lists the results for the DeathStarBench Memory dataset, where most strategies achieve comparable coverage. This is attributed to the relatively low variability in memory usage within the dataset. Consequently, KIMetrix proves more effective in scenarios involving high-variability metrics, where SREs derive greater practical benefit. Even in low variable scenarios, KIMetrix, however, achieves lower informative subset sizes.

\begin{figure}[ht]
    \begin{subfigure}{0.45\textwidth}
        \centering
        \includegraphics[width=\linewidth]{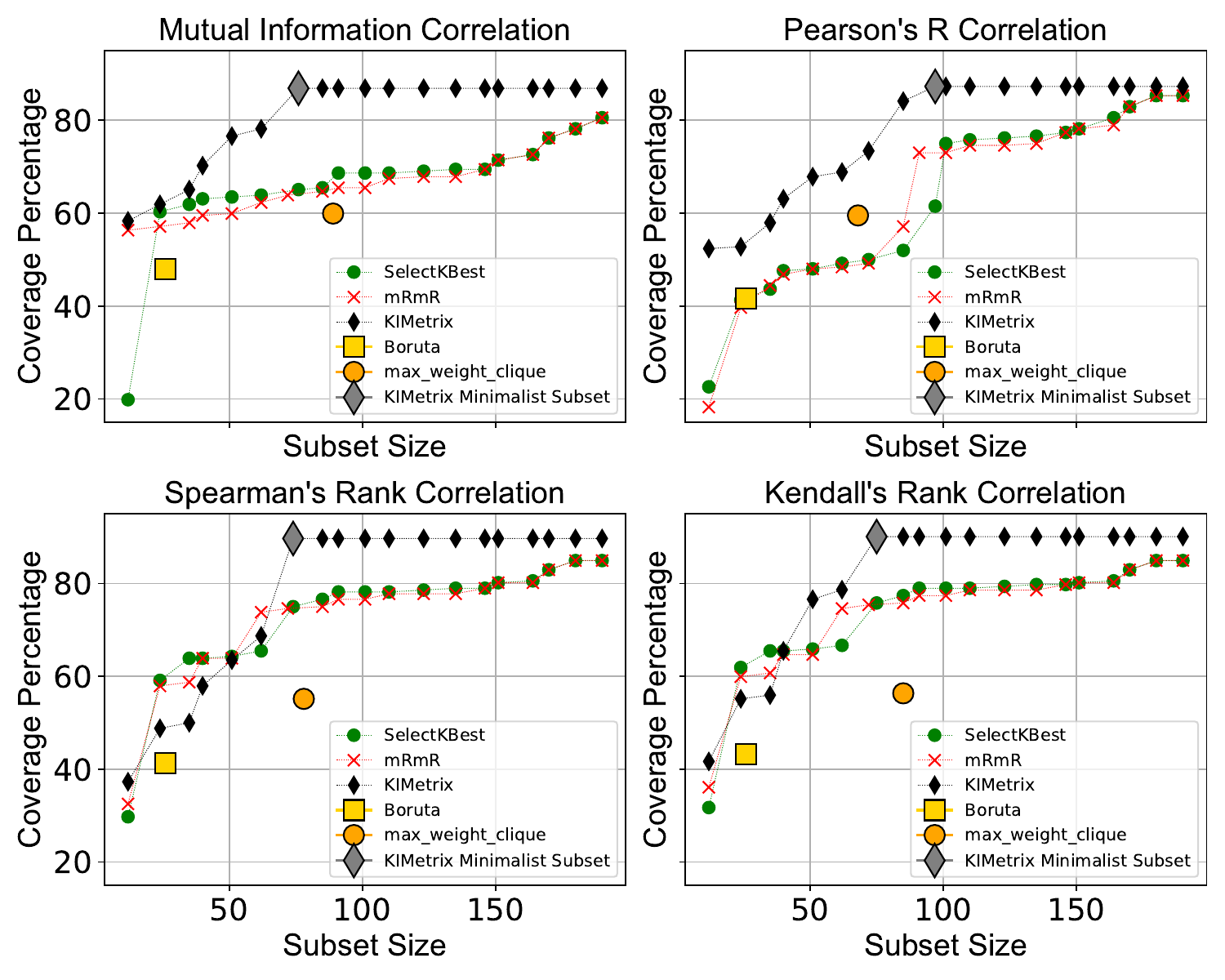} 
    \end{subfigure}
    \caption{Comparison of Feature Selection Methods for different subset sizes using KIMetrix \textit{logs} on \textit{Mix} data for $\mathcal{C}$}
    \label{fig:combined-baseline}
\end{figure}

In Figure \ref{fig:combined-baseline}, we plot the coverage percentage against subset sizes from KIMetrix’s AIMD execution logs. In addition to providing the informative metric subset, these logs capture the trade-off between each subset and its corresponding coverage. Unlike Boruta and Max Weighted Clique, which return a single subset size, KIMetrix dynamically offers multiple subsets, enhancing adaptability for SREs. The subset sizes derived from KIMetrix are used to feed into supervised approaches such as SelectKBest and mRMR for comparison purposes, where we compute the corresponding coverage for each size as demonstrated in Figure \ref{fig:combined-baseline}.\\

\begin{figure}
    \centering
    \begin{subfigure}[b]{0.98\linewidth}
        \centering
        \includegraphics[scale =0.24]{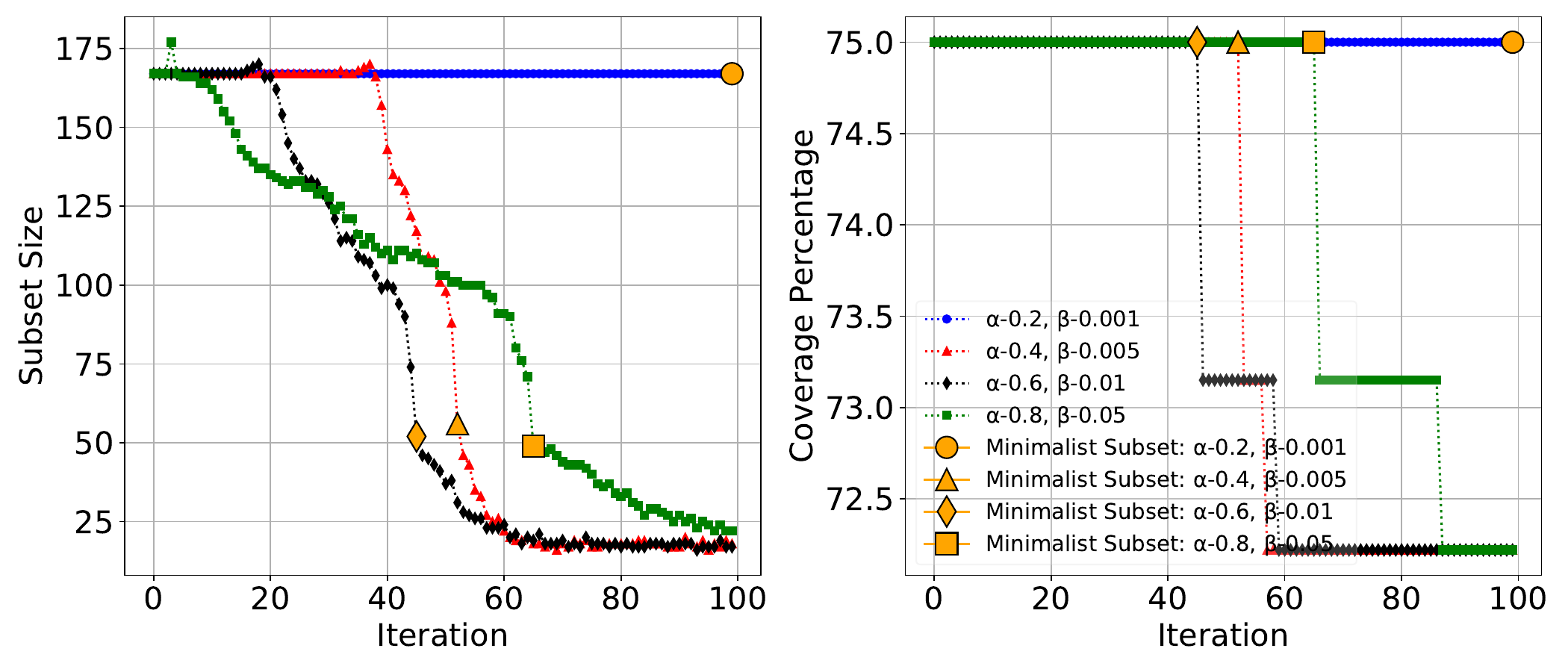}
        \caption{Healthy Data}
        \label{fig:aimd-plots-healthy}
    \end{subfigure}

    \begin{subfigure}[b]{0.98\linewidth}
        \centering
        \includegraphics[scale=0.24]{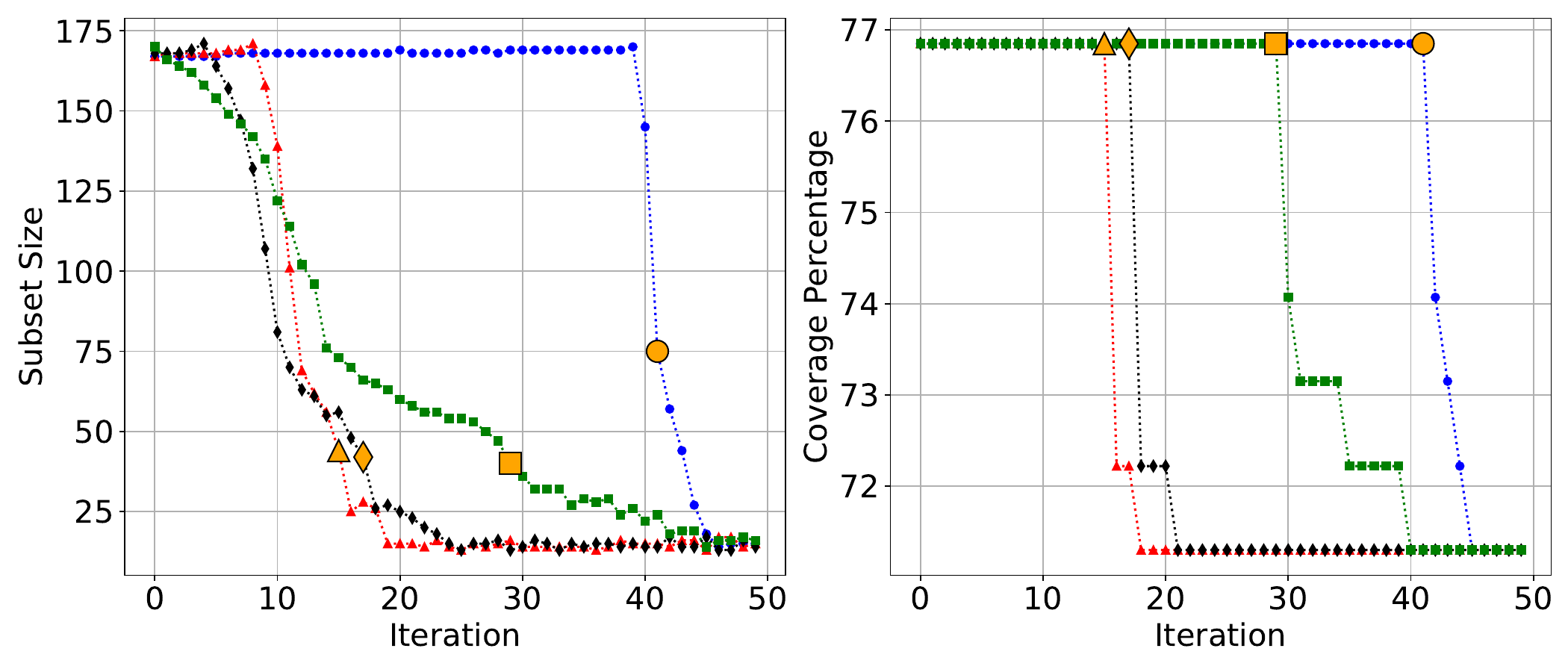}
        \caption{Mix Data}
        \label{fig:aimd-plots-mix}
    \end{subfigure}

    \caption{Trends observed for subset sizes and coverage with respect to the iterations on varying Algorithm 3's multiplicative and additive factors on Mix and Healthy data using Mutual Information Correlation.}
    \label{fig:aimd-plots}
\end{figure}

\noindent
\textbf{Q4: What impact does the AIMD parameters have?}
\noindent
We explore the impact of varying the AIMD parameters of Algorithm 3 on subset size and coverage, focusing specifically on the multiplicative ($\alpha$) and additive ($\beta$) factors in Figure~\ref{fig:aimd-plots}.
With the number of iterations set to 100 and an initial epsilon value of 0.5, we observe similar trends at 100 and 50 iterations for healthy and mix, respectively. This is due to the nature of the Healthy dataset, which exhibits greater metrics correlations. 
We now analyze the trends in two categories:
(1) Lower $\alpha$–$\beta$ factors:
With $\alpha = 0.2$, $\beta = 0.001$, subset growth is slow due to minimal $\epsilon$ updates, resulting in stable but limited coverage. In Figure~\ref{fig:aimd-plots-mix}, Mix data shows sharp declines when outside the tolerance range, as low $\alpha$ overly reduces $\epsilon$. Healthy data (Figure~\ref{fig:aimd-plots-healthy}) remains stable. Higher values ($\alpha = 0.4$, $\beta = 0.005$) lead to smoother growth and gentler declines.
(2) Higher $\alpha$–$\beta$ factors:
Larger $\beta$ (e.g., 0.01) drives faster $\epsilon$ growth, increasing subset size with minor dips, moderated by higher $\alpha$. With $\alpha = 0.8$ and $\beta = 0.05$, growth is faster and more stable.
Overall, orange markers in Figure~\ref{fig:aimd-plots} denote the final informative subset size per configuration. Higher $\alpha$–$\beta$ pairs yield faster convergence and smoother trends, though they may explore a narrower range of subsets.

\section{Related Work}
\label{sec:related_work}

\noindent
The field of cloud service monitoring that relies on defining alerts on a critical set of metrics has seen substantial focus. We discuss some recent approaches in this sphere.

Anomaly Detection: Initial explorations for time-series anomaly detection primarily utilized statistical techniques such as moving averages, standard deviation analysis, and z-score normalization \cite{Taylor2018}. Recent advancements have introduced neural network-based models specifically designed for time series anomaly detection \cite{Hundman2018, Fawaz2019}. 
In this context, KIMetrix's metrics can be used in conjunction with anomaly detection methodologies by invoking anomaly detectors with only the informative set, making the pipeline leaner.

Alerting Mechanisms: Alerting mechanisms typically rely on domain-specific, hand-crafted rules and include strategies like alert aggregation \cite{Chen2019}, alert correlation \cite{Chen2020a}, and alert ranking \cite{Chen2020b}. 
Incident detection and management emphasizes the swift identification and resolution of incidents \cite{li2021fighting, li2022intelligent}. For example, Li et al. \cite{li2021fighting} address the challenges of maintaining the availability of large-scale cloud computing platforms. 
In \cite{thalheim2017sieve}, the authors propose a mechanism to reduce the metric space via clustering, however do not consider the role of the microservice topology.
KIMetrix can yield a close look at the metric spectrum to aid SREs to define such alerts to aid incident management.

Service Level Objectives in Cloud Systems: Service Level Objectives (SLOs) is categorized into two primary areas. The first category defines SLOs from a cloud perspective and empirically examines the associated challenges \cite{jayathilaka2015response, mogul2017thinking, mogul2019nines}. 
The second category discusses models, algorithms, runtime mechanisms, and tools for SLO-native cloud resource management, ensuring performance guarantees for users \cite{nastic2020sloc, qiu2020firm}. While these works are pioneering in defining SLO concepts and challenges, they are neither targeted towards microservice applications nor do they deal with our objective of identifying the most critical metrics towards SLO definition.

\section{Conclusion and Future Work}
\label{sec:conclusion}

\noindent
In this paper, we presented KIMetrix, to aid SREs identify an informative set of metrics towards efficient observability monitoring for microservice-based applications using metrics and light-weight traces. By leveraging information-theoretic measures, it identifies the critical microservice-metric mappings in a topology-aware manner. Experimental evaluations show that KIMetrix presents SREs with significant insights into the metric-coverage trade-off that can be used to define potential alerts. Future avenues include incorporating logs in a lightweight manner, to further refine metric selection.

\bibliographystyle{IEEEtran}
\bibliography{references}

\begin{thebibliography}{10}
\providecommand{\url}[1]{#1}
\csname url@samestyle\endcsname
\providecommand{\newblock}{\relax}
\providecommand{\bibinfo}[2]{#2}
\providecommand{\BIBentrySTDinterwordspacing}{\spaceskip=0pt\relax}
\providecommand{\BIBentryALTinterwordstretchfactor}{4}
\providecommand{\BIBentryALTinterwordspacing}{\spaceskip=\fontdimen2\font plus
\BIBentryALTinterwordstretchfactor\fontdimen3\font minus
  \fontdimen4\font\relax}
\providecommand{\BIBforeignlanguage}[2]{{%
\expandafter\ifx\csname l@#1\endcsname\relax
\typeout{** WARNING: IEEEtran.bst: No hyphenation pattern has been}%
\typeout{** loaded for the language `#1'. Using the pattern for}%
\typeout{** the default language instead.}%
\else
\language=\csname l@#1\endcsname
\fi
#2}}
\providecommand{\BIBdecl}{\relax}
\BIBdecl

\bibitem{monitorassistant}
Z.~Yu, M.~Ma, C.~Zhang, S.~Qin, Y.~Kang, C.~Bansal, S.~Rajmohan, Y.~Dang,
  C.~Pei, D.~Pei \emph{et~al.}, ``Monitorassistant: Simplifying cloud service
  monitoring via large language models,'' in \emph{Companion Proceedings of the
  32nd ACM International Conference on the Foundations of Software
  Engineering}, 2024, pp. 38--49.

\bibitem{msr}
P.~Srinivas, F.~Husain, A.~Parayil, A.~Choure, C.~Bansal, and S.~Rajmohan,
  ``Intelligent monitoring framework for cloud services: A data-driven
  approach,'' in \emph{Proceedings of the 46th International Conference on
  Software Engineering: Software Engineering in Practice}, 2024, pp. 381--391.

\bibitem{thalheim2017sieve}
J.~Thalheim, A.~Rodrigues, I.~E. Akkus, P.~Bhatotia, R.~Chen, B.~Viswanath,
  L.~Jiao, and C.~Fetzer, ``Sieve: Actionable insights from monitored metrics
  in distributed systems,'' in \emph{Proceedings of the 18th ACM/IFIP/USENIX
  middleware conference}, 2017, pp. 14--27.

\bibitem{alerts4}
A.~Bhattacharyya, S.~A.~J. Jandaghi, S.~Sotiriadis, and C.~Amza, ``Semantic
  aware online detection of resource anomalies on the cloud,'' in \emph{2016
  IEEE International Conference on Cloud Computing Technology and Science
  (CloudCom)}.\hskip 1em plus 0.5em minus 0.4em\relax IEEE, 2016, pp. 134--143.

\bibitem{Chen2019}
J.~Chen, X.~He, Q.~Lin, H.~Zhang, D.~Hao, F.~Gao, Z.~Xu, Y.~Dang, and D.~Zhang,
  ``Continuous incident triage for large-scale online service systems,'' in
  \emph{2019 34th IEEE/ACM International Conference on Automated Software
  Engineering (ASE)}.\hskip 1em plus 0.5em minus 0.4em\relax IEEE, 2019, pp.
  364--375.

\bibitem{gan2019open}
Y.~Gan, Y.~Zhang, D.~Cheng, A.~Shetty, P.~Rathi, N.~Katarki, A.~Bruno, J.~Hu,
  B.~Ritchken, B.~Jackson \emph{et~al.}, ``An open-source benchmark suite for
  microservices and their hardware-software implications for cloud \& edge
  systems,'' in \emph{Proceedings of the Twenty-Fourth International Conference
  on Architectural Support for Programming Languages and Operating Systems},
  2019, pp. 3--18.

\bibitem{Chen2020}
J.~Chen, S.~Zhang, X.~He, Q.~Lin, H.~Zhang, D.~Hao, Y.~Kang, F.~Gao, Z.~Xu,
  Y.~Dang \emph{et~al.}, ``How incidental are the incidents? characterizing and
  prioritizing incidents for large-scale online service systems,'' in
  \emph{Proceedings of the 35th IEEE/ACM International Conference on Automated
  Software Engineering}, 2020, pp. 373--384.

\bibitem{cinque2022micro2vec}
M.~Cinque, R.~Della~Corte, and A.~Pecchia, ``Micro2vec: Anomaly detection in
  microservices systems by mining numeric representations of computer logs,''
  \emph{Journal of Network and Computer Applications}, vol. 208, p. 103515,
  2022.

\bibitem{panahandeh2024serviceanomaly}
M.~Panahandeh, A.~Hamou-Lhadj, M.~Hamdaqa, and J.~Miller, ``Serviceanomaly: An
  anomaly detection approach in microservices using distributed traces and
  profiling metrics,'' \emph{Journal of Systems and Software}, vol. 209, p.
  111917, 2024.

\bibitem{rouf2024instantops}
R.~Rouf, M.~Rasolroveicy, M.~Litoiu, S.~Nagar, P.~Mohapatra, P.~Gupta, and
  I.~Watts, ``Instantops: A joint approach to system failure prediction and
  root cause identification in microserivces cloud-native applications,'' in
  \emph{Proceedings of the 15th ACM/SPEC International Conference on
  Performance Engineering}, 2024, pp. 119--129.

\bibitem{RedHat2024}
\BIBentryALTinterwordspacing
{Red Hat}, ``{Red Hat OpenShift Enterprise Kubernetes Container Platform},''
  accessed: 2024-03-29. [Online]. Available:
  \url{https://www.redhat.com/en/technologies/cloud-computing/openshift}
\BIBentrySTDinterwordspacing

\bibitem{208870}
B.~Rabenstein and J.~Volz, ``Prometheus: A next-generation monitoring system
  (talk).''\hskip 1em plus 0.5em minus 0.4em\relax Dublin: {USENIX}
  Association, May 2015.

\bibitem{instana}
``Instana observability,'' https://instana.github.io/openapi/, 2024.

\bibitem{gamma}
\BIBentryALTinterwordspacing
G.~Somashekar, A.~Dutt, M.~Adak, T.~Lorido~Botran, and A.~Gandhi, ``Gamma:
  Graph neural network-based multi-bottleneck localization for microservices
  applications.'' in \emph{Proceedings of the ACM Web Conference 2024}, ser.
  WWW '24.\hskip 1em plus 0.5em minus 0.4em\relax New York, NY, USA:
  Association for Computing Machinery, 2024. [Online]. Available:
  \url{https://doi.org/10.1145/3589334.3645665}
\BIBentrySTDinterwordspacing

\bibitem{scikit-learn-selectkbest}
\BIBentryALTinterwordspacing
S.~learn developers. (2023) Selectkbest. [Online]. Available:
  \url{https://scikit-learn.org/stable/modules/generated/sklearn.feature\_selection.SelectKBest.html}
\BIBentrySTDinterwordspacing

\bibitem{mrmr}
H.~Peng, F.~Long, and C.~Ding, ``Feature selection based on mutual information
  criteria of max-dependency, max-relevance, and min-redundancy,'' \emph{IEEE
  Transactions on pattern analysis and machine intelligence}, vol.~27, no.~8,
  pp. 1226--1238, 2005.

\bibitem{boruta}
M.~B. Kursa and W.~R. Rudnicki, ``Feature selection with the boruta package,''
  \emph{Journal of statistical software}, vol.~36, pp. 1--13, 2010.

\bibitem{max-weighted-clique}
\BIBentryALTinterwordspacing
(2023) Maximum weighted clique. [Online]. Available:
  \url{https://networkx.org/documentation/stable/reference/algorithms/generated/networkx.algorithms.clique.max_weight_clique.html}
\BIBentrySTDinterwordspacing

\bibitem{Taylor2018}
S.~Taylor and B.~Letham, ``Forecasting at scale,'' \emph{The American
  Statistician}, vol.~72, no.~1, pp. 37--45, 2018.

\bibitem{Hundman2018}
K.~Hundman, V.~Constantinou, C.~Laporte, I.~Colwell, and T.~Söderström,
  ``Detecting spacecraft anomalies using lstms and nonparametric dynamic
  thresholding,'' in \emph{Proceedings of the 24th ACM SIGKDD International
  Conference on Knowledge Discovery \& Data Mining}.\hskip 1em plus 0.5em minus
  0.4em\relax ACM, 2018, pp. 387--395.

\bibitem{Fawaz2019}
H.~I. Fawaz, G.~Forestier, J.~Weber, L.~Idoumghar, and P.-A. Muller, ``Deep
  learning for time series classification: a review,'' \emph{Data mining and
  knowledge discovery}, vol.~33, no.~4, pp. 917--963, 2019.

\bibitem{Chen2020a}
Y.~Chen, X.~Yang, H.~Dong, X.~He, H.~Zhang, Q.~Lin, J.~Chen, P.~Zhao, Y.~Kang,
  F.~Gao \emph{et~al.}, ``Identifying linked incidents in large-scale online
  service systems,'' in \emph{Proceedings of the 28th ACM joint meeting on
  European software engineering conference and symposium on the foundations of
  software engineering}, 2020, pp. 304--314.

\bibitem{Chen2020b}
Z.~Chen, Y.~Kang, L.~Li, X.~Zhang, H.~Zhang, H.~Xu, Y.~Zhou, L.~Yang, J.~Sun,
  Z.~Xu \emph{et~al.}, ``Towards intelligent incident management: why we need
  it and how we make it,'' in \emph{ESEC/SIGSOFT FSE}.\hskip 1em plus 0.5em
  minus 0.4em\relax ACM, 2020, pp. 1487--1497.

\bibitem{li2021fighting}
L.~Li, X.~Zhang, X.~Zhao, H.~Zhang, Y.~Kang, P.~Zhao, B.~Qiao, S.~He, P.~Lee,
  J.~Sun \emph{et~al.}, ``Fighting the fog of war: Automated incident detection
  for cloud systems,'' in \emph{2021 USENIX Annual Technical Conference (USENIX
  ATC 21)}, 2021, pp. 131--146.

\bibitem{li2022intelligent}
Y.~Li, X.~Zhang, S.~He, Z.~Chen, Y.~Kang, J.~Liu, L.~Li, Y.~Dang, F.~Gao, Z.~Xu
  \emph{et~al.}, ``An intelligent framework for timely, accurate, and
  comprehensive cloud incident detection,'' \emph{ACM SIGOPS Operating Systems
  Review}, vol.~56, no.~1, pp. 1--7, 2022.

\bibitem{jayathilaka2015response}
H.~Jayathilaka, C.~Krintz, and R.~Wolski, ``Response time service level
  agreements for cloud-hosted web applications,'' in \emph{Proceedings of the
  Sixth ACM Symposium on Cloud Computing}, 2015, pp. 315--328.

\bibitem{mogul2017thinking}
J.~C. Mogul, R.~Isaacs, and B.~Welch, ``Thinking about availability in large
  service infrastructures,'' in \emph{Proceedings of the 16th Workshop on Hot
  Topics in Operating Systems}, 2017, pp. 12--17.

\bibitem{mogul2019nines}
J.~C. Mogul and J.~Wilkes, ``Nines are not enough: Meaningful metrics for
  clouds,'' in \emph{Proceedings of the Workshop on Hot Topics in Operating
  Systems}, 2019, pp. 136--141.

\bibitem{nastic2020sloc}
S.~Nastic, A.~Morichetta, T.~Pusztai, S.~Dustdar, X.~Ding, D.~Vij, and
  Y.~Xiong, ``Sloc: Service level objectives for next generation cloud
  computing,'' vol.~24, no.~3.\hskip 1em plus 0.5em minus 0.4em\relax IEEE,
  2020, pp. 39--50.

\bibitem{qiu2020firm}
H.~Qiu, S.~S. Banerjee, S.~Jha, Z.~T. Kalbarczyk, and R.~K. Iyer,
  ``$\{$FIRM$\}$: An intelligent fine-grained resource management framework for
  $\{$SLO-Oriented$\}$ microservices,'' in \emph{14th USENIX symposium on
  operating systems design and implementation (OSDI 20)}, 2020, pp. 805--825.

\end{thebibliography}

\end{document}